\documentclass[proof]{WileyASNA-v1}

\articletype{Article Type}%

\received{17 March 2024}
\revised{}
\accepted{19 May 2024}

\begin{document}

\title{Size of the accretion disc in the recurrent nova T~CrB}

\author[1]{R. K. Zamanov}

\author[1]{K. A. Stoyanov}

\author[1]{V. Marchev}

\author[1]{M. Minev}

\author[2]{D. Marchev}

\author[1]{M. Moyseev}

\author[3]{J. Mart\'i}

\author[4,5]{M. F. Bode}

\author[1]{R. Konstantinova-Antova}

\author[1]{S. Stefanov}

\authormark{Zamanov, Stoyanov, Marchev \textsc{et al}}

\address[1]{\orgname{Institute of Astronomy and National Astronomical Observatory, 
                     Bulgarian Academy of Sciences}, 
            \orgaddress{72 Tsarigradsko Shose, Bulgaria}}

\address[2]{\orgdiv{Department of Physics and Astronomy}, 
            \orgname{Shumen University "Episkop Konstantin Preslavski"},  
            \orgaddress{115 Universitetska Str., 9700 Shumen, Bulgaria, \country{Bulgaria}}}

\address[3]{\orgdiv{Departamento de F\'isica, Escuela Polit\'ecnica Superior de Ja\'en}, 
            \orgname{Universidad de Ja\'en}, 
	    \orgaddress{Campus Las Lagunillas s/n, A3-420, 23071, Ja\'en, \country{Spain}}}

\address[4]{\orgname{Astrophysics Research Institute, Liverpool John Moores University}, 
            \orgaddress{IC2, 149 Brownlow Hill, Liverpool, L3 5RF, \country{UK}}}

\address[5]{\orgdiv{Office of the Vice Chancellor}, 
            \orgname{Botswana International University of Science and Technology}, 
	    \orgaddress{Private Bag 16, Palapye, \country{Botswana}}}

\corres{*R. Zamanov \email{rkz@astro.bas.bg}}


\abstract{We present high resolution (0.06 \AA~px$^{-1}$) spectroscopic observations
of the recurrent nova T Coronae Borealis
obtained during the last 1.5 years (September 2022 -- January 2024), 
with the 2.0m RCC telecope of 
the Rozhen National Astronomical Observatory, Bulgaria. 
Double-peaked emission is visible in the $H_\alpha$ line
after the end of the superactive state. 
We subtract the red giant contribution and measure the distance between the peaks ($\Delta v_a$)
of the line. For the period 
July 2023 -- January 2024, we find that $\Delta v_a$ is in range  
$90 < \Delta v_a < 120$~km~s$^{-1}$. 
Assuming that the emission is from the accretion disc around the white dwarf,
we find average radius of the accretion disc $R_{disc} = 89 \pm 19$~R$_\odot$,
which is approximately equal to the Roche lobe size of the white dwarf.
Our results indicate that tidal torque plays an important role but  
that the disc  can extend up to the Roche lobe of the accreting star. } 

\maketitle


\section{Introduction}
  
T CrB (HD 143454, NOVA CrB 1946, NOVA CrB 1866) is a famous recurrent nova
having recorded eruptions in 1866, in 1946 
and possibly in 1217 and  1787 (Schaefer 2023a).
A new outburst can be expected in the near future 
(Luna et el. 2020; Maslennikova et al. 2024; Schaefer 2023b),
which will make T~CrB the brightest nova outburst since  Nova 1500 Cyg in 1975.

The  nature of the binary system  T~CrB was revealed when 
(i) Sanford (1949) 
discovered that the radial velocity of the M giant 
varies with a period of 230.5 days; 
(ii) Peel (1985) and Lines et al. (1988) found 
that the red giant is ellipsoidally shaped;
and 
(iii) Selvelli et al. (1992) using $IUE$ spectra identified that the hot component of the system 
is an accreting white dwarf.
T~CrB is a member of a small group of symbiotic recurrent novae with only six confirmed members -- RS~Oph, T~CrB, V3890~Sgr, V745~Sco, LMC~S154 and V618~Sgr  (Ilkiewicz et al. 2019; Merc et al. 2023).

In our previous paper (Zamanov et al. 2023a) 
we analysed UBV photometry to investigate the evolution 
of the hot component of T CrB through the superactive state and linked this to the proposed upcoming recurrent nova outburst.
Here we analyse optical spectroscopic observations and estimate the accretion disc radius.

 \begin{figure}   
 \vspace{8.9cm} 
  \includegraphics{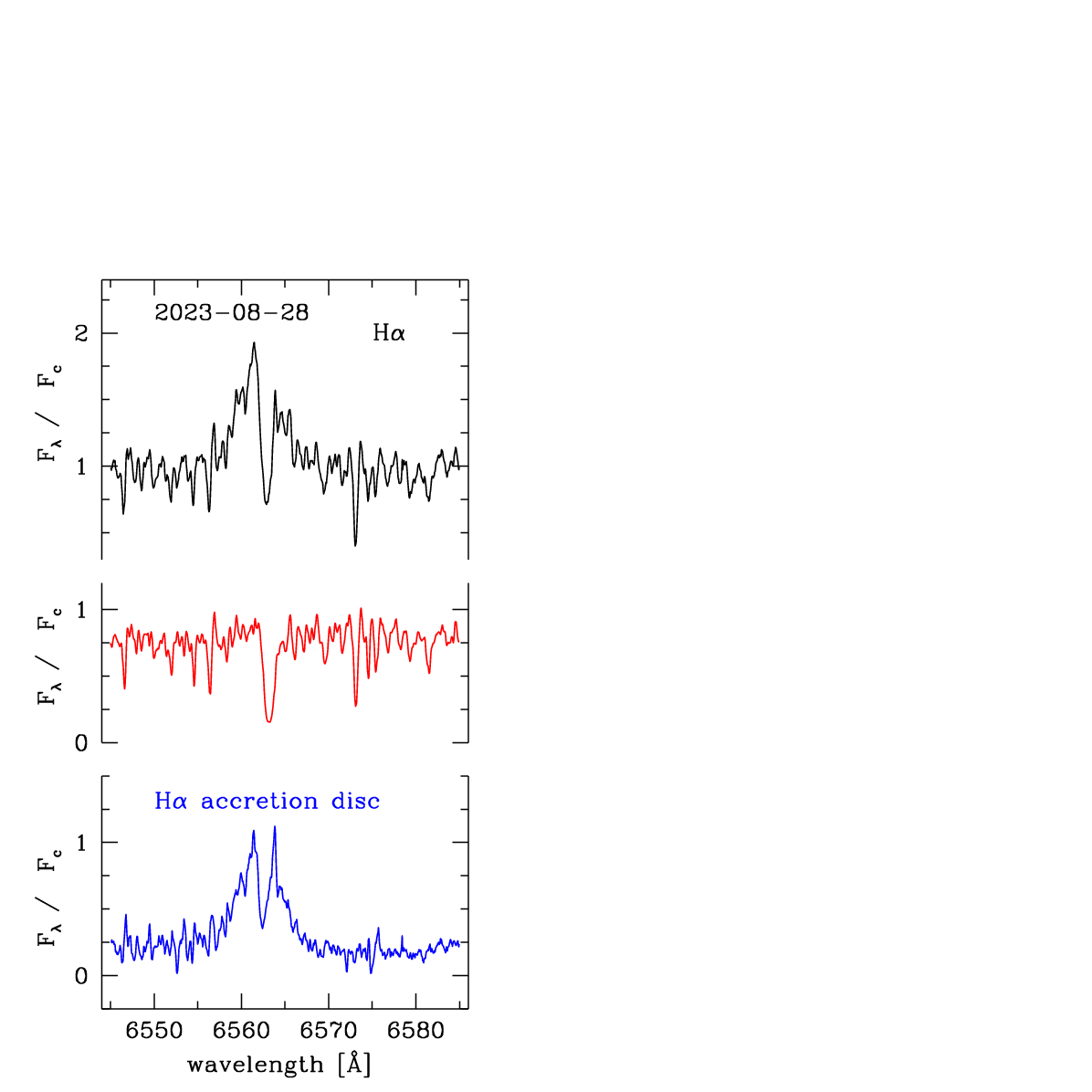}	
  \caption[]{H$\alpha$ emission line of T~CrB observed on 2023-08-28 (black). 
             The subtraction of the red giant spectrum (red) reveals 
             a double-peaked emission line (blue). 
	     See text for further details.  }
  \label{f.RG}
 \end{figure}

 \begin{figure*}     
  \vspace{11.5cm}   
  \includegraphics{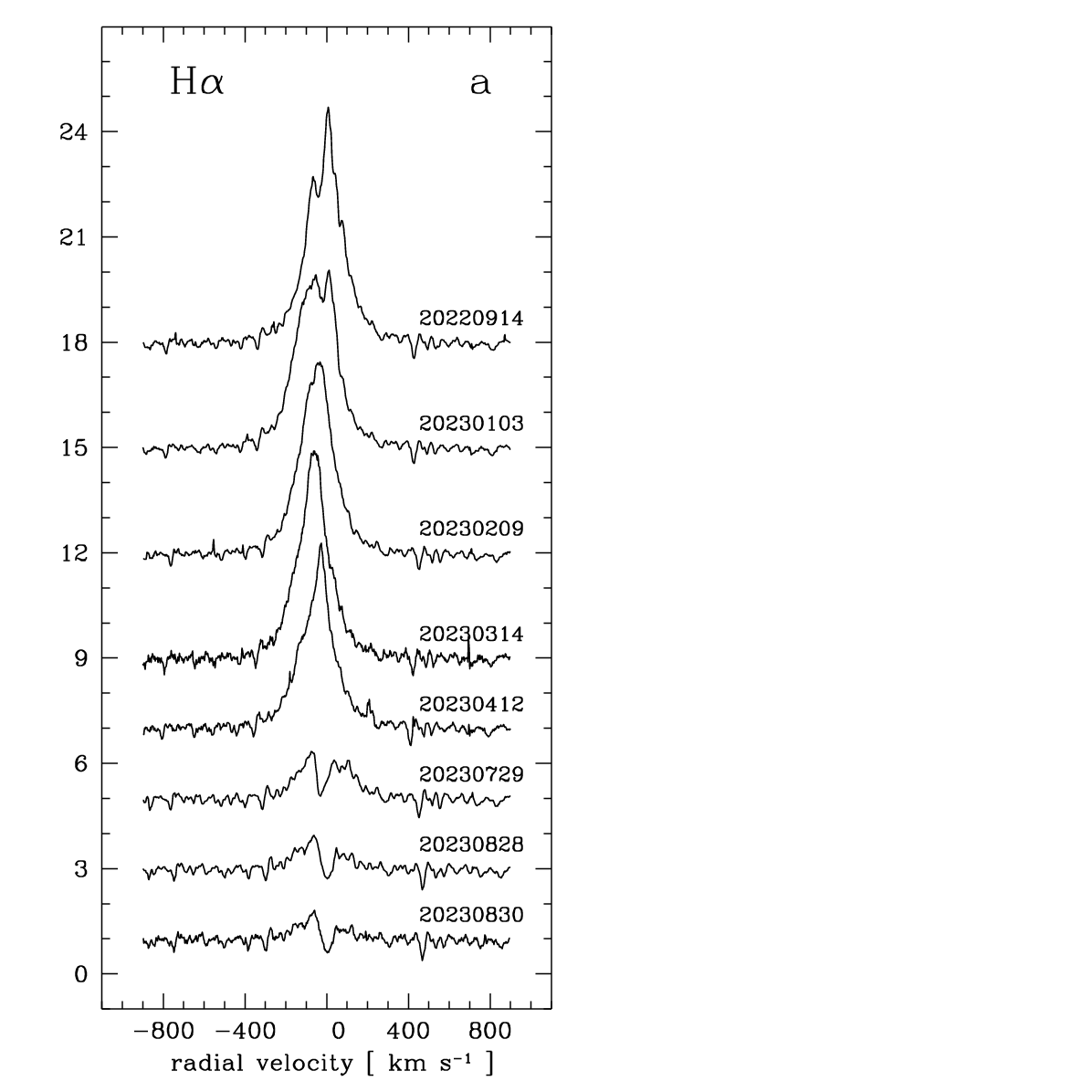} 
  \includegraphics{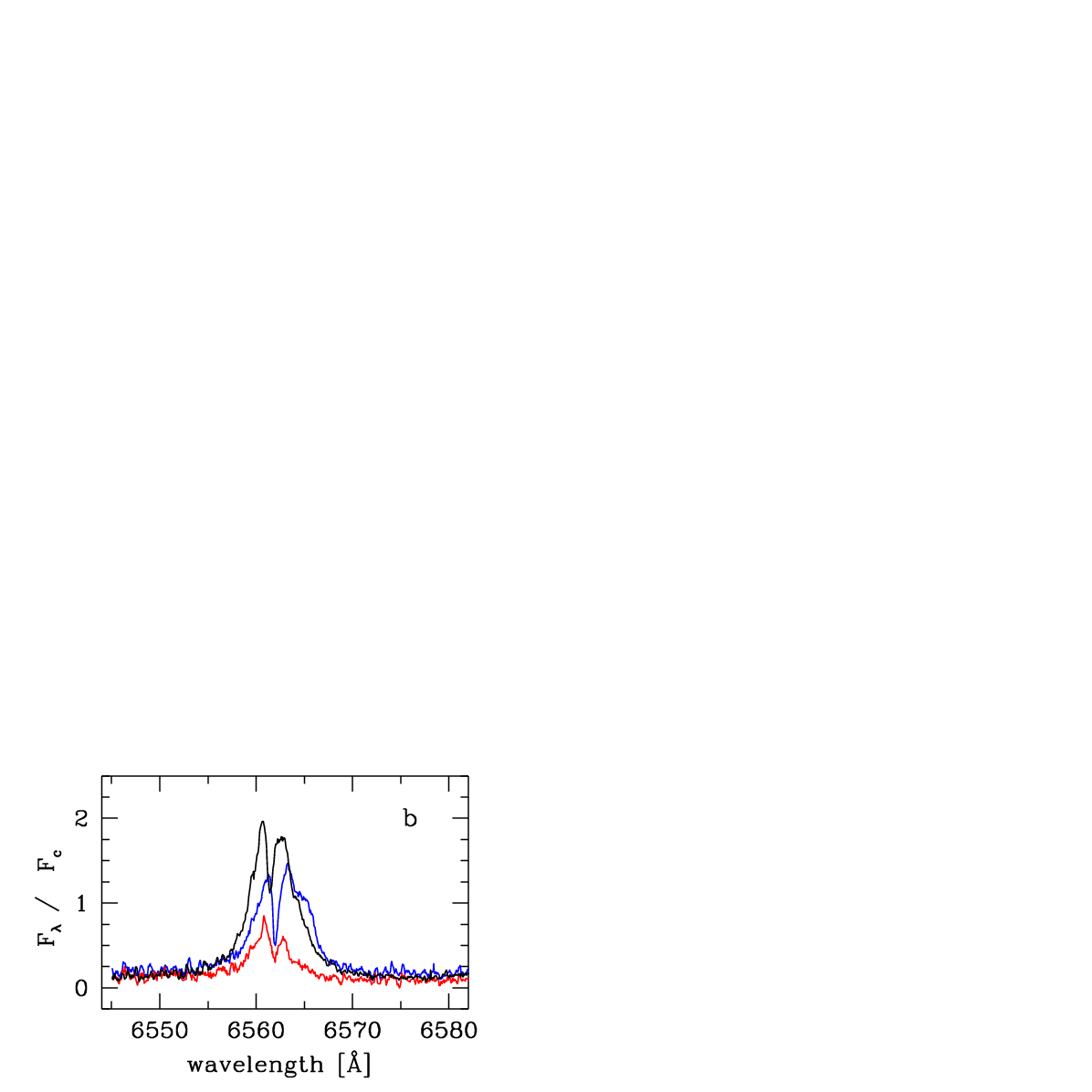}	   
  \caption[]{Variability of H$\alpha$  emission line profile of T~CrB, {\bf a -} the  observed 
             $H\alpha$ emission, {\bf b -} a few examples of the  $H\alpha$ emission 
	     of the accretion disc (after the red giant subtraction). 
	     The colours are:  blue (2023-07-29), red (2023-10-23), and black (2024-01-23). }
  \label{f.Hab}      
\end{figure*}	     

 \begin{figure}     
  \vspace{11.0cm}   
  \includegraphics{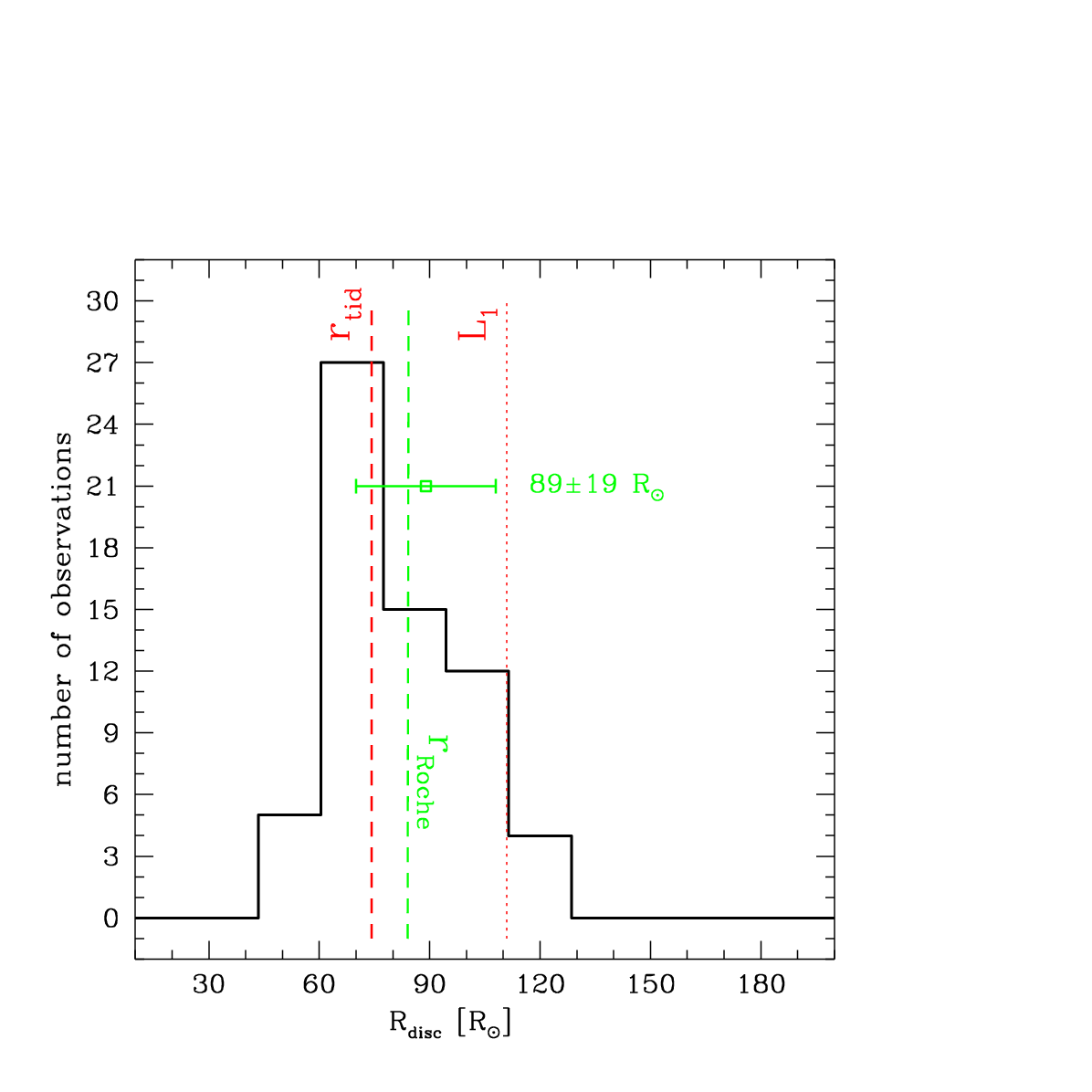} 
  \caption[]{Histogram of the radius of the H$\alpha$-emitting disc of 
  $\;$ T~CrB.   The vertical lines indicate the Roche Lobe (green dashed line),  
  $L_1$ (red dotted line) and $r_{tid}$ (red dashed line).
  The average size of the disc ($R_d =89 \pm 19$~R$_\odot$) is also marked.  
  }
  \label{f.hi}      
\end{figure}	     


\section{Observations}

73 optical spectra of T~CrB on 21 nights were secured with the 
ESpeRo Echelle spectrograph (Bonev et al. 2017)
on the 2.0~m~RCC telescope in the Rozhen  National Astronomical Observatory, Bulgaria. 
On each spectrum we measure the equivalent width of the H$\alpha$ line.
The typical error is $\pm 5$\%. 
We then subtract the spectrum of HD134807, 
which is the red giant used in Stanishev et al. (2004). 
An example of the subtraction of the red giant is shown in Fig.$\,$\ref{f.RG}$\!\!\!.$
The double-peaked nature of the line is visible before the subtraction on most of the spectra,  
however the subtraction reveals this more clearly 
and gives us the possibility 
to measure accurately the separation of the peaks.  

%

The variability of the H$\alpha$ emission line
of T~CrB is presented in Fig.$\,$\ref{f.Hab}$\!\!\!a.$ 
The spectra are normalized 
to the local continuum and a constant is added to each spectrum.
In Fig.$\,$\ref{f.Hab}$\!\!\!b$ are plotted three 
spectra after the subtraction of the red giant contribution.
Double-peaked $H\alpha$ emission
coming from the hot component is clearly visible. 

The spectroscopic observations of T~CrB are summarized in Table~\ref{t.1}$\!\!\!.$ 
In the table are given the start of the observation
(in the format YYYY-MM-DD HH:MM), number of exposures and exposure time in minutes, 
the equivalent width of H$\alpha$ line, $EW(H_\alpha)$,
the distance between the peaks of H$\alpha$, $\Delta v_a$, 
and the calculated disc size (see also Sect.\ref{s.Huang}). 
In Table~\ref{t.1} are given the average values for each night.
In Table~\ref{t.2} are given the measurements 
of $\Delta v_a$ for each spectrum. 
The distance between the peaks is measured in a way identical to 
that in Zamanov et al. (2023b).  

\begin{table*}
  \centering
  \caption{ Spectroscopic observations of T~CrB. 
   }
   \begin{tabular}{lcrrrcccccr} 
   \hline
date-obs      & phase        & exp.time & EW(H$\alpha$) &  $\Delta v_a$    & $R_{disc}$   &  \\
              &              &  [min]   &    [\AA]       &  [km~s$^{-1}$]  &  [$R_\odot$] &  \\
 \\
 2022-09-14 17:41 & 0.365 & 60	 &  28.2  &  \\
 2023-01-04 02:57 & 0.855 & 60	 &  28.4  &  \\
 2023-01-07 03:00 & 0.868 & 60	 &  30.4  &  \\
 2023-02-10 00:58 & 0.017 & 60	 &  24.9  &  \\
 2023-03-14 22:53 & 0.162 & 3x15 &  24.8  &  \\
 2023-03-29 21:58 & 0.228 & 60	 &  28.8  &  \\  
 2023-04-12 21:02 & 0.289 & 60	 &  20.9  &  \\
 2023-06-07 20:23 & 0.535 & 60	 &  10.9  &  \\
 2023-07-29 19:35 & 0.763 & 9x15  &  6.8   & $103.7 \pm 4  $  & $83.2 \pm 6$   & \\
 2023-07-30 18:48 & 0.767 & 6x15  &  7.2   & $102.1 \pm 2  $  & $85.6 \pm 4$   & \\
 2023-08-28 19:11 & 0.895 & 7x15  &  3.0   & $117.8 \pm 3  $  & $64.4 \pm 4$   & \\
 2023-08-29 18:37 & 0.899 & 10x15 &  3.2   & $118.4 \pm 2  $  & $63.7 \pm 3$   & \\
 2023-08-30 17:51 & 0.904 & 13x15 &  2.1   & $118.9 \pm 3  $  & $63.2 \pm 3$   & \\
 2023-08-31 19:31 & 0.908 & 3x15  &  2.4   & $120.1 \pm 3  $  & $61.9 \pm 3$   & \\ 
 2023-10-23 16:37 & 0.140 & 2x30  &  1.5   & $89.7  \pm 2  $  & $110.8\pm 4$   & \\
 2023-12-25 02:51 & 0.415 & 2x45  &  3.5   & $96.6  \pm 3  $  & $95.6 \pm 4$   & \\
 2023-12-26 02:44 & 0.419 & 4x25  &  6.4   & $89.9  \pm 5  $  & $111.1\pm 8$   & \\
 2023-12-27 02:45 & 0.424 & 1x50  &  4.1   & $90.2  \pm 7  $  & $109.7\pm 8$   & \\
 2023-12-28 02:44 & 0.428 & 1x60  &  4.2   & $98.9  \pm 7  $  & $91.2 \pm 8$   & \\
 2024-01-22 03:00 & 0.542 & 1x30  &  7.3   & $94.8  \pm 7  $  & $99.2 \pm 8$   & \\
 2024-01-23 01:52 & 0.542 & 6x20  &  8.9   & $90.9  \pm 2  $  & $108.0\pm 3$   & \\ 
 \hline 	  	     					
 \end{tabular}  						    
\label{t.1}				  
\end{table*}					  

\section{Results}
  


From 2016 until March 2023, T CrB was in a superactive state
(Munari et al. 2016; Munari 2023) characterised
by an increase in the mean brightness and the appearance of high-ionization emission lines (HeII4686,
[OIII]4959, 5007, [NeIII]3869, etc.) as well as  a prominent soft 
X-ray component (Zhekov \& Tomov 2019). 
Our spectroscopic observations are obtained during the last year of the 
superactive state and after it. 

During the period September 2022 -- April 2023, the H$\alpha$ emission 
is strong with $EW(H\alpha)$ in the range from $20$~\AA\ to  $30 $~\AA. 
This is due to the brighter hot component, an increased ionization in the companion wind, and most of  the H$\alpha$  emission
coming not from the accretion disc, but from the ionized wind of the red giant
(Munari 2023).
This also is visible in the radio observations, which indicate that during
the superactive state 
T~CrB displays higher emission in the radio, consistent with optically thick thermal 
bremsstrahlung emission from a photoionized source, 
and an increased ionization in the companion wind, 
driven by high accretion rate (Linford et al. 2019; Zamanov et al. 2023a). 
After the end of the superactive state  the equivalent width of $H_\alpha$ decreased 
to $EW(H\alpha) \approx  11$~\AA\  in June 2023, 
$EW(H\alpha) \approx 7$~\AA\  in July 2023,  
and  $EW(H\alpha) \approx 2$~\AA\ in August-October 2023. 
A double-peaked emission profile is visible in our observations obtained 
during the period July 2023 --  January 2024. 
The minimum of the H$\alpha$ emission is in August-October 2023, 
when $EW(H\alpha) \approx 2$~\AA.  


Hereafter we adopt for T~CrB an orbital period   227.5687~d,  
(Fekel et al. 2000), 
mass of the white dwarf $1.37 \pm 0.13$~$M_\odot$, 
mass of the red giant  $1.12 \pm 0.23$~$M_\odot$,
inclination $i=67^o.5$ (Stanishev et al. 2004)
and  zero eccentricity (Kenyon \& Garcia 1986). 

\subsection{Disc size}
\label{s.Huang}

The subtraction of the red giant contribution 
reveals that on all the spectra in the period July 2023 -- January 2024, 
the H$\alpha$  emission line of the hot component displays a double-peaked profile
(see Fig.~\ref{f.RG}$\!\!\!).$ 
The two peaks have almost equal intensity. 
This is an indication that H$\alpha$ is formed in a disc (e.g. Horne \& Marsh 1986).
We assume that this is a Keplerian accretion disc which surrounds  the white dwarf.  

For  emission lines coming from a Keplerian disc,
the peak separation ($\Delta v$) can be regarded as a measure of 
the outer radius ($R_{disc}$) of the emitting disc (Huang 1972): 
 \begin{equation}
   \Delta v = 2 \; \sin{i} \; \sqrt{GM_{wd}/R_{disc} \:}, 
  \label{H1}
  \end{equation}
where $G$ is the gravitational constant,  
$M_{wd}$ is the mass of the white dwarf and 
$i$ is the inclination angle of the disc axis to the line of sight.

For our July 2023 -- January 2024 observations (see Table~\ref{t.1}$\!\!\!)$, we measure
$90 < \Delta v_\alpha < 120$~km~s$^{-1}$, 
with average value $\Delta v_\alpha  = 102 \pm 12$~km~s$^{-1}$.   
From Eq.\ref{H1} we estimate the size of the H$\alpha$ 
emitting disc as $62 \le  R_{disc}  \le 111$~$R_\odot$, 
with average value $R_{disc} = 89 \pm 19$~$R_\odot$.
The average values are calculated using the values given in Table~\ref{t.1}$\!\!\!.$  
A histogram of the distribution (based on the data in Table~\ref{t.2}$\!\!\!)$
is presented in Fig.~\ref{f.hi}$\!\!\!.$ 
The minimum of disc size is in August 2023, 
when we estimate  $R_{disc} \approx 63$~R$_\odot$ (see Table~\ref{t.1}$\!\!\!).$

\subsection{Roche lobe size}
\label{s.RL}

With the binary parameters and the Kepler's third law
we calculate the distance between the components of T~CrB 
$a=212.6$~$R_\odot$.
We estimate, that the inner Lagrangian point, $L_1$, 
is located at a distance 111 R$_\odot$ from the white dwarf. 
The Roche lobe radius of the accreting star is given by the formula 
(Eggleton 1983):
\begin{equation}
r_{Roche} / a = (0.49 q^{2/3})/[ 0.6 q^{2/3} + \ln (1 + q^{1/3})],
\end{equation}
where $q=M_1/M_2$ is the mass ratio. 
Using this formula and mass ratio $q  = 1.22$, 
we estimate $r_{Roche} / a = 0.396$ and   
Roche lobe radius of the white dwarf 84.3~$R_\odot$, which means that the average 
size of the accretion disc in July 2023 -- January 2024
is approximately equal to the size of the Roche lobe of the white dwarf.

Another important important parameter is the tidal radius, which
depends on the mass ratio and is described 
with a polynomial fit (Smak 2020):
\begin{equation}
\begin{split}
r_{tid}/r_{Roche}=0.830 + 0.860 p - 4.974 p^2  \\
     + 12.410 p^3 - 14.842 p^4 + 6.903p^5, 
\end{split}
\end{equation}
where $p=M_2/M_1$. 
According to this formula for T~CrB we estimate  $r_{tid}/r_{Roche} =0.880$ and
$r_{tid} = 74.2$~R$_\odot$. 

In Fig.~\ref{f.hi}$\!\!\!$, as well as the histogram of the calculated  values of $R_{disc}$ 
are plotted three vertical lines:  the Roche lobe radius ($r_{Roche}$, blue dashed line),  
the distance of $L_1$ from the white dwarf (red dotted line) and $r_{tid}$ (red dashed line). 
The average disc size  $R_{disc}$ is marked in green  
and coincides with the radius of the Roche lobe around the white dwarf.

%
%

\section{Discussion}

The mass donor in T~CrB is a red giant and the system is classified as a symbiotic star. 
The symbiotic stars are long-period interacting binary systems composed of a hot component, 
a cool giant and a nebula formed from material lost by the donor star 
and ionized by the radiation of the hot component (Mikolajewska 2012). 
Their orbital periods vary from a few hundred days up to 100 years. 
Depending on the orbital period and the distance between the components,  
the  accretion onto the white dwarf can occur through gravitational
capture of the slow stellar wind from the giant component,
Bondi-Hoyle-Littleton accretion (Bondi \& Hoyle 1944), 
wind Roche-lobe overflow, WRLOF (Mohamed \& Podsiadlowski 2012), 
or a Roche lobe overflow when the system is semidetached.  

In many S-type symbiotic stars (e.g. RW~Hya, SY~Mus, AR~Pav, 
YY~Her, CI~Cyg, BF~Cyg) with P$_{orb}<900$~d,  
modulations are detected in the light curves with half-orbital period.
Such modulations are visible in the optical/near-IR bands 
and are due to ellipsoidal variations 
of the mass donor (Mikolajewska 2003; Yudin et al. 2005; Rutkowski et al. 2007). 
This is also the case in T CrB, where the ellipsoidal variations of the red giant 
are clearly  visible in B, V, R, I bands when the hot component is in a low 
state (Munari et al. 2016). 
The mass transfer from a tidally distorted red giant filling the Roche Lobe  
is expected to form an accretion disc similar to that of the cataclysmic variables.
Indeed, the behaviour of T CrB suggests that it is effectively a dwarf nova with an extremely
long orbital period, closely related to SU~UMa dwarf novae (Ilkiewicz et al. 2023). 

Duschl (1986) estimated that the outer radius of the accretion disc in symbiotic systems 
is in the range 15 -- 55 $R_\odot$.  
Leedjarv et al. (1994) 
calculated the outer radius of the accretion disc of the symbiotic star 
CH~Cyg  $\sim$ 66~R$_{\odot}$. 
Robinson et al. (1994) found acceptable fits to high resolution spectra for 3 symbiotic stars on the assumption that the double-peaked line profiles arose from accretion discs, 
where for T CrB the outer disc radius was estimated as $\approx 37$~R$_\odot$; 
for CH Cyg $\approx 120$~R$_\odot$; and for AG Dra $\approx 19$~R$_\odot$.
However, these authors note that it was highly unlikely that the lines in each case were emitted solely from a disc, and they did not for example subtract any contribution from the red giant in each system.  
Our estimates for $R_{disc}$ lie within the general range of the estimates above.
 

Smak (2020) suggests that the outer radius of the disc 
in the case of Roche lobe overflow
is controlled by  tidal torques, which prevent
the disc from expanding beyond the tidal radius. 
In Fig.~\ref{f.hi}$\!\!$ there is a well defined peak in the distribution of $R_{disc}$. 
The tendency for the disc size to cluster at a specific level is related
to the truncation of the disc at specific disc radii (e.g. Coe et al. 2006).
Not surprisingly the peak of the histogram 
corresponds to $R_{disc} \approx r_{tid}$ (Fig.~\ref{f.hi}$\!\!\!),$ 
in agreement with the suggestions of Smak (2020), 
that  tidal torque is an important factor in binaries with Roche lobe overflow.
However our results also indicate that it does not
prevent the disc extending to the Roche lobe 
and even in some cases to the inner Lagrangian point  $L_1$. 
The fact that we measure an average value $R_{disc} \approx r_{Roche}$ indicates
that the limiting factor for the disc size in T~CrB is the Roche lobe.
This suggests that in such cases there is  probably no accretion stream from 
$L_1$ and the matter flow through $L_1$ enters almost immediately into the
outer parts of the accretion disc.

\section{Conclusions}\label{sec5}
From September 2022 to January 2024, we observed spectroscopically the recurrent nova T~CrB.
During the latter months (July 2023 -- January 2024),  double-peaked $H\alpha$ 
emission is visible. We performed 73 measurements of the distance between the peaks of 
H$\alpha$ and estimate at each point the radius of the accretion disc 
thought to give rise to the line emission.  
We find that it is in the range
$60 < R_{disc} < 120$~R$_\odot$, with average $R_{disc} = 89 \pm 19$~R$_\odot$, 
which is  $\approx r_{Roche}$. 
Our results indicate the tidal torque plays a role, 
however the key limiting factor for the disc size is the Roche lobe. 

\section*{Acknowledgments}
We acknowledge the anonymous referee for the useful
comments.  The research infrastructure is supported by the
\fundingAgency{Ministry of Education and Science of Bulgaria}
(\fundingNumber{Bulgarian National Roadmap for Research Infrastructure}).
This work is partly supported by the 
\fundingAgency{Spanish Ministerio de Ciencia e Innovacion},
Agencia Estatal de Investigacion (Ref. \fundingNumber{PID2022-136828NB-C42}).  
D.M. acknowledges support by project RD-08-137/2024 from Shumen University Science Fund.



\subsection*{Conflict of interest}

The authors declare no potential conflict of interests.


{}



\section*{Appendix}

Here is given Table~\ref{t.2} containing the measurements of the distance 
between the peaks of the H$\alpha$ line
and the calculated disc radius. 

\begin{table*}
  \centering
  \caption{T~CrB -- radius of the H$\alpha$ emitting disc.
  In the columns are given the start of the exposure (in format \\
  YYYY-MM-DD~HH:SS), 
  exposure time in minutes, the measured distance between 
  the peaks of the H$\alpha$ line,  
  and the calculated radius of the H$\alpha$ emitting disc. 
  }
   \begin{tabular}{ccrlccccccr} 
   \hline
date-obs & exp-time & $\Delta v_a$  & $R_{disc}$  & date-obs & expos. & $\Delta v_a$      & $R_{disc}$  & \\
        & [min]   & [km s$^{-1}$] & [$R_\odot$] &          & [min]   & [km s$^{-1}$]  & [$R_\odot$] & \\
  2023-07-29 19:35 & 15 &  95.5  &  97.7 &     2023-08-31 19:31 & 15 & 123.1  &  58.8 & \\
  2023-07-29 19:50 & 15 & 103.7  &  82.9 &     2023-08-31 19:47 & 15 & 119.2  &  62.8 & \\
  2023-07-29 20:06 & 15 & 103.8  &  82.8 &     2023-08-31 20:03 & 15 & 118.1  &  64.0 & \\
  2023-07-29 20:22 & 15 & 103.7  &  83.0 &     2023-10-23 16:37 & 30 & 90.6   & 108.8 & \\
  2023-07-29 20:38 & 15 & 104.8  &  81.3 &     2023-10-23 17:08 & 30 & 88.9   & 112.8 & \\  
  2023-07-29 20:53 & 15 & 102.9  &  84.3 &     2023-12-25 02:51 & 45 & 95.7   & 97.5  & \\
  2023-07-29 21:09 & 15 & 105.3  &  80.5 &     2023-12-25 03:37 & 45 & 97.6   & 93.7  & \\
  2023-07-29 21:24 & 15 & 108.1  &  76.3 &     2023-12-26 02:44 & 25 & 96.8   &  95.3 & \\
  2023-07-29 21:40 & 15 & 105.4  &  80.4 &     2023-12-26 03:09 & 25 & 90.9   & 107.9 & \\
  2023-07-30 18:48 & 15 & 104.4  &  81.9 &     2023-12-26 03:36 & 25 & 86.4   & 119.4 & \\
  2023-07-30 19:04 & 15 & 101.4  &  86.7 &     2023-12-26 04:02 & 25 & 85.6   & 121.8 & \\
  2023-07-30 19:20 & 15 & 102.3  &  85.2 &     2023-12-27 02:45 & 50 & 90.2   & 109.7 & \\
  2023-07-30 19:38 & 15 & 103.6  &  83.1 &     2023-12-28 02:44 & 60 & 98.9   &  91.2 & \\
  2023-07-30 19:54 & 15 & 99.2   &  90.7 &     2024-01-22 03:00 & 30 & 94.8   &  99.2 & \\
  2023-07-30 20:18 & 15 & 101.9  &  85.9 &     2024-01-23 01:52 & 20 & 89.7   & 110.9 & \\
  2023-08-28 19:11 & 15 & 115.1  &  67.3 &     2024-01-23 02:13 & 20 & 89.3   & 111.8 & \\
  2023-08-28 19:27 & 15 & 116.3  &  65.9 &     2024-01-23 02:34 & 20 & 91.7   & 106.2 & \\
  2023-08-28 19:44 & 15 & 122.3  &  59.7 &     2024-01-23 02:54 & 20 & 92.1   & 105.2 & \\
  2023-08-28 19:59 & 15 & 118.0  &  64.1 &     2024-01-23 03:15 & 20 & 90.6   & 108.8 & \\
  2023-08-28 20:15 & 15 & 114.3  &  68.3 &     2024-01-23 03:36 & 20 & 92.0   & 105.4 & \\ 
  2023-08-28 20:31 & 15 & 121.5  &  60.5 & \\
  2023-08-28 20:48 & 15 & 116.8  &  65.3 & \\
  2023-08-29 18:37 & 15 & 117.9  &  64.1 & \\
  2023-08-29 18:53 & 15 & 117.8  &  64.3 & \\
  2023-08-29 19:09 & 15 & 119.1  &  62.9 & \\
  2023-08-29 19:24 & 15 & 122.0  &  59.9 & \\
  2023-08-29 19:40 & 15 & 114.8  &  67.7 & \\
  2023-08-29 19:58 & 15 & 119.8  &  62.2 & \\
  2023-08-29 20:13 & 15 & 118.3  &  63.7 & \\
  2023-08-29 20:29 & 15 & 118.8  &  63.2 & \\
  2023-08-29 20:46 & 15 & 118.3  &  63.8 & \\			      
  2023-08-29 21:02 & 15 & 116.9  &  65.2 & \\
  2023-08-30 17:51 & 15 & 119.7  &  62.3 & \\
  2023-08-30 18:07 & 15 & 121.3  &  60.7 & \\
  2023-08-30 18:22 & 15 & 120.9  &  61.0 & \\
  2023-08-30 18:38 & 15 & 115.2  &  67.2 & \\
  2023-08-30 18:53 & 15 & 115.1  &  67.4 & \\
  2023-08-30 19:09 & 15 & 120.5  &  61.4 & \\
  2023-08-30 19:24 & 15 & 115.5  &  66.9 & \\
  2023-08-30 19:40 & 15 & 117.4  &  64.7 & \\
  2023-08-30 19:55 & 15 & 119.7  &  62.3 & \\
  2023-08-30 20:11 & 15 & 124.3  &  57.7 & \\
  2023-08-30 20:27 & 15 & 118.3  &  63.7 & \\
\hline 			        		      
\end{tabular}  		        			     
\label{t.2}			           
\end{table*}			        	   

\end{document}